\documentclass[]{raa}

\usepackage{graphicx,times}             
\usepackage{latexsym}

\begin{document}

   \title{XMM-Newton observation of the eclipsing binary Algol
}

 \volnopage{ {\bf 2009} Vol.\ {\bf 9} No. {\bf XX}, 000--000}
   \setcounter{page}{1}

   \author{Xue-Juan Yang
      \inst{1,2}
   \and Fang-Jun Lu
      \inst{2}
   \and B. Aschenbach
      \inst{3}
   \and Li Chen
      \inst{4}
   }

   \institute{Faculty of Materials, Optoelectronics and Physics, Xiangtan University,
             Xiangtan 411105, China; {\it xjyang@xtu.edu.cn}\\
        \and
             Particle Astrophysics Center, Institute of High Energy Physics,
Chinese Academy of Sciences, Beijing 100049, China
        \and
             Max-Planck-Institut f\"ur Extraterrestrische Physik, 85741
Garching, Germany
        \and
             Department of Astronomy, Beijing Normal University, Beijing
100875, China\\
\vs \no
   {\small Received [year] [month] [day]; accepted [year] [month] [day] }
}

   \abstract{
   We present an {\sl XMM-Newton} observation
   of the eclipsing binary Algol which contains an X-ray dark
   B8V primary and an X-ray bright K2IV secondary.
   The observation covered the optical secondary eclipse and
   captured an X-ray flare that was eclipsed by the B star.
   The EPIC and RGS spectra of Algol in its quiescent state
   are described by a two-temperature plasma model.
   The cool component has a temperature around 6.4$\times$10$^{6}$ K
   while that of the hot component ranges from 2 to 4.0$\times$10$^{7}$ K.
   Coronal abundances of C, N, O, Ne, Mg, Si and Fe were obtained for
   each component for both the quiescent and the flare phases, with generally
   upper limits for S and Ar, and C, N, and O for the hot component. F-tests
   show that the abundances need not to be different between the cool and the
   hot component and between the quiescent and the flare phase with the exception
   of Fe. Whereas the Fe abundance of the cool component remains constant at
   $\sim$0.14, the hot component shows an Fe abundance of $\sim$0.28, which increases
   to $\sim$0.44 during the flare. This increase is expected from the chromospheric
   evaporation model. The absorbing column density $N_H$ of the quiescent emission
   is 2.5$\times10^{20}$ cm$^{-2}$, while that of the flare-only emission is
   significantly lower and consistent with the column density of the interstellar medium.
   This observation substantiates earlier suggestions of the presence of
   X-ray absorbing material in the Algol system.
   }

   \authorrunning{X.-J. Yang et al.}            
   \titlerunning{{\sl XMM-Newton} observation of Algol}  
   \maketitle

%

\section{Introduction}

Algol ($\beta$ Per) is a nearby eclipsing binary with an early
main-sequence (B8V) primary and a cool subgiant (K2IV)
secondary. This system is only 28.46 pc away (Hipparcos parallax
measurements, ESA 1997) and has a period of about 2.87 d and an
orbit inclination of $81^{\circ}$. The radii of the two companion
stars ($R_{B}$ and $R_K$) are 3.0 and 3.4 $R_{\odot}$, respectively,
and their separation is
14.14 $R_{\odot}$ (Hill et al. 1971; Richards 1993).
It is a semi-detached system with the K star filling its
Roche lobe. The mass transfer occurs in the form of gas streams from the K
star to the B star and ends up in an annulus around the B star. (Richards
1993; Richards et al. 1995; Richards 2001; Richards 2004;
Retter et al. 2005).
Algol was first detected in X-rays by {\sl SAS} 3 (Schnopper et al.
1976) and confirmed as a strong X-ray emitter by Harnden et al. (1977).
It is generally accepted that the K star has an
active corona and accounts for most of the X-ray flux of the
system, whereas the B star is X-ray dark (White et al. 1980). This was
proved for the first time by Chung et al.
(2004) by  the detection of Doppler
shifts of the spectra caused by orbital motion of the K star. Given an
active corona, flares appear typically every day with a duration of several
to many hours.

Algol shows elemental abundances which differ from those of the sun (Antunes et al. 1994;
Schmitt \& Ness 2004). Antunes et al. (1994) quote abundances of Fe, O, Mg,
Si, S, Ar and Ca which are lower than the solar photospheric values by a factor of 2-3,
and N to be less than 0.1. With the exception of N these results have been generally confirmed by
{\sl Chandra} observations (Schmitt \& Ness 2004). Schmitt \& Ness (2002)
studied a sample of late type stars including Algol and they found an enhancement of
N. Drake (2003) also suggested that the N abundance
is enhanced by a factor of 3, and C is depleted  by a factor
of 10 (both relative to HR1009, whose C/N abundance is consistent with that
of the solar photosphere). These C and N abundances would lead to the conclusion
that the K star has lost at least half of its initial mass.

That flares have different elemental abundances has been concluded from
previous missions, such as {\sl GINGA}
(Stern et al. 1992), {\sl ROSAT} (Ottmann \& Schmitt 1996), {\sl BeppoSAX}
(Favata \& Schmitt 1999), {\sl Chandra} {Nordon \& Behar 2007} and
{\sl XMM-Newton} {Nordon \& Behar 2008}. They suggest that the elemental abundances
 of Algol's flaring region resemble more the photospheric than the
quiescent coronal abundances, which might be a consequence of chromospheric
evaporation (G\"{u}del et al. 1999; Favata \& Micela 2003
and references therein). At the beginning of the event, fresh
chromospheric material is evaporated in
the flaring loop(s), and enhances the corona elemental abundances; once the
material has been transported to the flaring corona structure, the
fractionation mechanism, which is responsible for the lower
abundance, would start operating, bringing the corona abundance
back to its quiescent value.

The absorbing column density ($N_H$) inferred  from X-ray
observations is usually higher than the column density of the
interstellar medium (ISM) between Algol and the observer.
Welsh et al. (1990) gave an upper limit of the ISM column density
of 2.5$\times10^{18}$ cm$^{-2}$ using the ISM Na{\small I} D line
absorption of the B8 primary. Stern et al. (1995) got a similar
result from the He/H ratio using EUVE observations.
The $N_H$ derived from the {\sl ROSAT} PSPC observation of
Algol is about 5$\times10^{19}$ cm$^{-2}$ (Ottmann 1994).
Favata \& Schmitt (1999), using {\sl BeppoSAX}, reported that $N_H$ increased
 up to $>10^{21}$ cm$^{-2}$ during the flare
rise and decreased during the decay, which may possibly be associated
with moving, cool material in the line of sight, e.g. a major
coronal mass ejection.

In this paper we present a time resolved spectroscopy of Algol, using {\sl XMM-
Newton} European Photon Imaging Camera (EPIC) and Reflection Grating
Spectrometer (RGS) observations of Algol. Compared with the instruments used in
earlier studies the EPIC cameras are the first ones to have a
combination of moderate spectral resolution, wide energy coverage,
and high sensitivity, while RGS provides us with unparalleled
sensitivity for high resolution spectroscopy in the soft X-ray
band. These advantages permit us to perform  a more detailed
diagnosis of the plasma properties and their evolution as far as the corona of
the K type secondary is concerned. The {\sl XMM-Newton}
observation covered the secondary eclipse and an X-ray flare was detected.
Schmitt et al. (2003) presented a detailed study of the geometry of the flare,
and we analyze the X-ray spectra of Algol in both the quiescent and flaring
states. We can therefore study the mechanism of the X-ray emission
of Algol in both the quiescent and the flaring states. In $\S$ 2,
we describe the observation and data reduction. The spectral properties
of the overall and flare-only emissions are given in $\S$ 3. The proposed local
cool absorbing gas structures are discussed in $\S$ 4 and a summary is provided
in $\S$ 5.

\section{Observation and Data Reduction}

Algol was observed with {\sl XMM-Newton} on 2002-02-12 from 04:02:39 to
18:45:09 UTC. We use data from the EPIC-PN (Str\"uder et al.
2001), the (two) EPIC-MOS (Turner et al. 2001) detectors and the
(two) RGS spectrometers (den Herder et al. 2001). The EPIC-PN was
turned on at 04:42:18  and turned off at 18:34:45 UTC, with a net
exposure time of about 50 ks, while the two EPIC-MOS and the two RGS
detectors had a bit longer exposure, from 04:09:01 to 18:39:28 UTC
and from 04:02:39 to 18:42:59, respectively. The EPIC-PN  has an
energy coverage of 0.15-15 keV, energy resolution of 80 eV  at 1
keV, and a time resolution of 73 ms in the full-frame imaging mode
in which the detector was operated for this observation. The EPIC-MOS
has a narrower
energy coverage of 0.15-12 keV, a better energy resolution of 70
eV at 1 keV, and a time resolution of 2.6 s in the standard
full-frame mode that was used in this observation. The RGS
allows much better energy resolution measurements (i.e. 2.9 eV at
1 keV) in the soft X-ray range (0.3 to 2.1 keV). Algol was positioned
at the aim point, and thick filters were used in order to reduce
the effect of strong optical loading of the CCD chips.

The X-ray data were analyzed using the XMMSAS software package. We
found that there is no significant background flare and thus no
time interval except the observation gaps had to be eliminated
from the dataset. As Algol is very bright, the center part of the
source has been affected seriously by pile-up for the EPIC data, but
not for the RGS data. Therefore, in the EPIC data analysis, we used the PN
counts extracted from an annulus of 21.75 arcsec inner radius and
125 arcsec outer radius. For the MOS inner and
outer radii of 21.25 and 125 arcsec were taken for the selection annulus,
respectively. According to the integrated PSFs of EPIC-PN and EPIC-MOS at 1.5
keV (which is near the peak of Algol's emission),
photons in the annuli chosen make up about 10\% of the total
source photons for the EPIC-PN and EPIC-MOS detectors (Aschenbach
et al. 2000). This factor was taken into account when we
calculated the fluxes and emission measures.

The corresponding EPIC PN and MOS1 0.3 to 10 keV  and the RGS1 0.3 to 2.1 keV
background corrected light curves for the selected annuli and the entire
observation are shown in Fig. 1 with time
bins of 100 s each. Using the ephemeris $Primary~minimum = HJD
2,445,739.0030+2.^d8673285E$ (Al-Naimiy et al. 1985; Kim et al. 1989), the
observation falls in the phase range 0.36-0.56 which includes the optical
secondary eclipse.

In order to study the temporal evolution of the spectrum, we divided the
observation into 6 segments and created a spectrum for each.
The integration time is 5000 to 10000 s for these spectra.
For the EPIC-PN, we only used the
single events (PATTERN=0) for the spectral analysis, and the
background was extracted from an annulus of 490 and 635
arcsec (the area contaminated by out-of-time events was
excluded). For the two EPIC-MOS, we used both single and double
events, and the background spectrum was extracted from an annulus
of 490 and 660 arcsec. For the two RGS only the
first order data were used. Each spectrum was fitted with a
two-temperature thermal plasma model (VMEKAL model; Mewe et al.
1995) in XSPEC. The free parameters are temperatures (T1, T2),
emission measures (EM1, EM2) and abundances of C, N, O, Ne, Mg, Si, S,
Ar and Fe for each of the two components. The abundances are in units
of solar abundances given by Anders \& Grevesse (1989). The
WABS model (Morrison \& McCammon 1983) was used to take care of
interstellar photo-electric absorption.  In the
joint fit of the EPIC and RGS spectra, the RGS1, RGS2 and the EPIC
emission measures were allowed to be different, because the
EPIC counts were selected from a restricted annulus as described above,
and the corresponding emission measures were of course much lower than those of
the two RGS.

To study the properties of the flaring plasma, the flare
period was cut in two sections, i.e. sections
\# 4 \& 5 (c.f. Fig. 1). We chose the quiescent spectrum
of the first 8000 s of the observation (c.f. Fig. 1) as
background spectrum, which is justified by the fact that the
X-ray eclipse often starts around
phase 0.40 (e.g., Schmitt \& Favata 1999), corresponding to
about 10000 s after exposure start of the present observation.
Given that the quiescent emission dominates the RGS energy band,
use of the RGS data could not be made because of the fairly low signal
to noise ratio. Taking into account that the quiescent emission was
also eclipsed during the flare, we scaled the background spectrum
by a factor of 0.9, which is the eclipsed fraction of the
quiescent emission (Schmitt \& Favata 1999). It should be noted
that we assume that the quiescent emission
arises from an isotropic corona of the K star
and remains unchanged during the flare,
and thus no rotation modulation was taken into account when we
dealt with the background emission of the flare-only spectrum.
The spectral analysis was performed with XSPEC. A one-temperature
MEKAL model in combination with the WABS model was used.

\begin{figure*}
\includegraphics[width=12cm,clip]{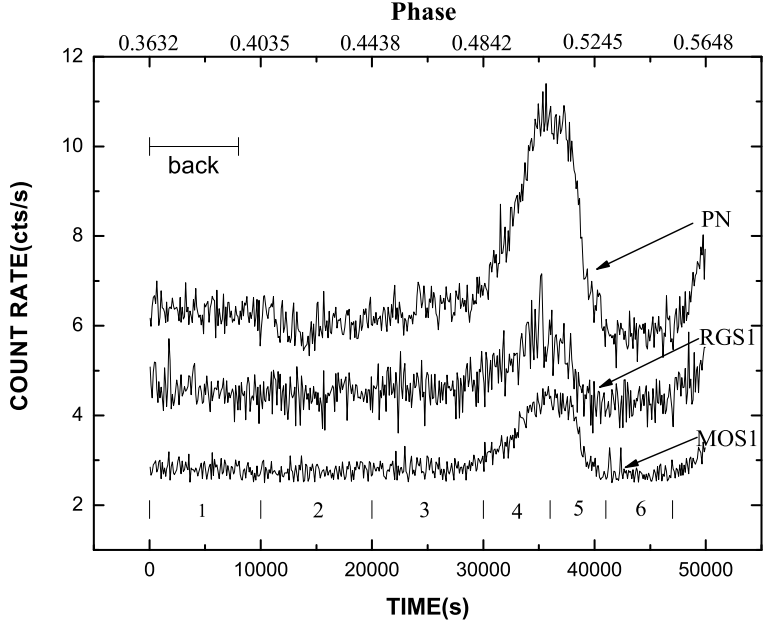}
\caption{EPIC PN and MOS1 0.3 to 10 keV and RGS1
0.3 to 2.1 keV light curves
 of Algol as a function of time and phase respectively, with time bins
of 100 s. The PN and MOS1 curves have each been moved up by 1 cts/s and the RGS1 up
by 2 cts/s for clarity.  The horizontal line segment
marked as ``back'' defines the time interval during which the background
spectrum was taken for the analysis of the  flare-only
spectra.}
\end{figure*}

\section{Spectra}
\subsection{The spectra of the overall X-ray emission}
The best-fit model parameters of the 6 spectra as well as their
errors are listed in Table 1 \& 2  and plotted in Fig. 2, 3 \& 4.
The errors are calculated in XSPEC taking into account the full error
projection, but they represent just the statistical errors at 90\% confidence.
As an example, Fig. 5 shows the model fit to the spectrum \#4.

From Tables 1 \& 2 and Fig. 2, 3 \& 4, we derive the following results:

(I) The best-fit  $N_H$ values are about $\sim 2.5\times10^{20}$
cm$^{-2}$, and remain nearly constant during the entire observation.
Considering that the potential effect of the EPIC cameras on the $N_H$
values might be due  their low efficiency in the soft energy band,
we also fit the EPIC and RGS data separately. The $N_H$ from the EPIC data is
typically $1.6\pm0.5 \times10^{20}$ cm$^{-2}$, while $3\pm0.4 \times10^{20}$
cm$^{-2}$ from the RGS data. The results given in  Table 1 \& Fig 2 correspond to
the
weighted mean value of the two. Within the error bars the  $N_H$ appears to be constant in time.

This $N_H$ is a factor of 100 higher than the upper limit of
the ISM H{\small I} + H$_2$ column density towards Algol derived
by Welsh et al. (1990). Fitting the spectra with
$N_H$ fixed at 2.5$\times 10^{18}$ cm$^{-2}$ ends up with significantly higher $\chi^2$ values.
With these two  sets of $N_H$, $\chi^2$ and degrees of freedom, we can carry out
an F-test using the {\sl ftest}-routine in XSPEC. The results show that
most of the spectra have of a probability of
$<0.01\%$ with $N_H$ that small. An $N_H$ value increase during the
flare rise, which has been detected with BeppoSAX (Favata \& Schmitt 1999)
has not been detected in our observation, i.e., there is no evidence for major mass
ejection during the flare to the extent that extra absorption occurs.
We also tried to fit each of the two spectral components
with different $N_H$ values but the fit did not improve significantly.
i.e., the $N_H$ values of the two components show no significant
difference.

(II) Both the temperature and the flux of the hot component show
significant temporal variations similar to the light curve.
This confirms the {\sl ROSAT} PSPC results
that the increase of the  count rate during the flare is predominantly
the result of an emission measure increase of the hot component
(Ottmann 1994; Ottmann \& Schmitt 1996).

(III) The RGS spectra are very similar to those from the {\sl
Chandra} LETGS (Ness et al. 2002). The emission lines of N, O, Ne, Mg and Fe
are clearly visible as shown in Fig. 5(b). The Ly$\alpha$ lines from Mg{\small XII}
(8.42 \AA), Ne {\small X}(12.14 \AA), O{\small VIII}(18.79 \AA), and N{\small VI}(24.78 \AA) can
easily be recognized and have similar relative strengths as in the LETGS spectra.
One exception is that the Si emission lines are absent in the RGS spectra. We speculate
that this is due to the low detecting efficiency of RGS around 6 \AA. We note that the RGS
spectra in the quiescent and flare state are very similar to each other. This is not
surprising since the RGS spectra are dominated by the cool component, while most of the flare
emission is from the hot component.

\begin{table*}
\caption[]{Spectral properties of Algol (2T VMEKAL model)}
\label{obslog}
\scriptsize{
\begin{tabular}{ccccccccccc}
\noalign{\smallskip} \hline\hline \noalign{\smallskip}
No.&$N_H$&T1&EM1$\ast$&flux1$\dagger$&T2&EM2$\ast$&flux2$\dagger$&$\chi^2_{\sl red}$&DOF$\ddagger$&F-test$\ast$$\ast$\\
&($\frac {10^{20}}{cm^2}$)&($10^6$K)&($\frac{10^{53}}{cm^3}$)&($\frac {10^{-11}erg}{{cm^2}s}$)
&($10^6$K)&($\frac {10^{53}}{cm^3}$)&($\frac {10^{-11}erg}{{cm^2}s}$)&&&probability\\
\noalign{\smallskip} \hline \noalign{\smallskip}

1   &$  2.51    _{  -0.24   }^{ +0.44   }$&$    6.27    _{  -0.12   }^{ +0.58   }$&$    6.55    _{  -0.60   }^{ +1.49   }$& 5.14    &$  20.31   _{  -0.58   }^{ +1.74   }$&$    9.76    _{  -2.30   }^{ +0.09   }$& 8.43    &   1.62    &   1979    &   $3.28\times10^{-4}$ \\
2   &$  2.35    _{  -0.34   }^{ +0.22   }$&$    6.38    _{  -0.12   }^{ +0.23   }$&$    6.28    _{  -0.12   }^{ +1.77   }$& 5.15    &$  21.82   _{  -0.35   }^{ +1.16   }$&$    8.70    _{  -1.25   }^{ +0.31   }$& 8.05    &   1.65    &   2000    &   $9.26\times10^{-5}$ \\
3   &$  2.55    _{  -0.39   }^{ +0.17   }$&$    6.50    _{  -0.00   }^{ +0.35   }$&$    6.50    _{  -0.13   }^{ +2.26   }$& 5.24    &$  22.17   _{  -0.12   }^{ +1.74   }$&$    9.57    _{  -1.04   }^{ +0.09   }$& 8.93    &   1.63    &   2051    &   $9.50\times10^{-4}$ \\
4   &$  2.25    _{  -0.35   }^{ +0.23   }$&$    6.96    _{  -0.12   }^{ +0.23   }$&$    9.34    _{  -1.00   }^{ +2.39   }$& 6.67    &$  37.14   _{  -1.39   }^{ +1.39   }$&$    12.48   _{  -0.75   }^{ +0.09   }$& 16.1    &   1.53    &   1326    &   $1.30\times10^{-13}$ \\
5   &$  2.19    _{  -0.49   }^{ +0.21   }$&$    6.73    _{  -0.23   }^{ +0.23   }$&$    8.20    _{  -0.98   }^{ +1.65   }$& 5.89    &$  33.54   _{  -1.16   }^{ +1.28   }$&$    12.48   _{  -0.84   }^{ +0.28   }$& 15.5    &   1.50    &   1136    &   $2.14\times10^{-7}$ \\
6   &$  2.81    _{  -0.40   }^{ +0.43   }$&$    5.80    _{  -0.23   }^{ +0.35   }$&$    4.59    _{  -1.01   }^{ +0.61   }$& 4.49    &$  20.66   _{  -1.16   }^{ +0.12   }$&$    9.13    _{  -0.28   }^{ +0.82   }$& 8.18    &   1.71    &   957     &   $4.29\times10^{-2}$\\

\noalign{\smallskip} \hline \noalign{\smallskip}
\end{tabular}
}

$\ast$ EPIC emission measure\\
$\ast$$\ast$ F-test probability for adopting identical abundances for the two components.\\
$\dagger$ unabsorbed EPIC energy flux in range of 0.5-10 keV.\\
$\ddagger$ degrees of freedom.
\end{table*}

\begin{table*}
\caption[]{Spectral properties of Algol (2T VMEKAL model). The top six rows are for the {\sl cool} component
and the lower six rows are for the {\sl hot} component.}
\label{obslog}
\scriptsize{
\begin{tabular}{cccccccccc}
\noalign{\smallskip} \hline\hline \noalign{\smallskip}
No.&C&N&O&Ne&Mg&Si&S&Ar&Fe\\
\noalign{\smallskip} \hline \noalign{\smallskip}

1   &$  0.11    _{  -0.11   }^{ +0.11   }$&$    2.18    _{  -1.23   }^{ +0.01   }$&$    0.26    _{  -0.08   }^{ +0.01   }$&$    1.04    _{  -0.42   }^{ +0.03   }$&$    0.27    _{  -0.10   }^{ +0.03   }$&$    0.30    _{  -0.13   }^{ +0.06   }$&$    <0.08   $&$    <0.64   $&$    0.15    _{  -0.03   }^{ +0.03   }$\\
2   &$  0.05    _{  -0.05   }^{ +0.10   }$&$    1.93    _{  -0.71   }^{ +0.23   }$&$    0.23    _{  -0.05   }^{ +0.03   }$&$    0.72    _{  -0.17   }^{ +0.09   }$&$    0.26    _{  -0.05   }^{ +0.08   }$&$    0.24    _{  -0.10   }^{ +0.09   }$&$    <0.08   $&$    <0.39   $&$    0.15    _{  -0.02   }^{ +0.01   }$\\
3   &$  0.09    _{  -0.07   }^{ +0.10   }$&$    1.49    _{  -0.62   }^{ +0.13   }$&$    0.21    _{  -0.04   }^{ +0.08   }$&$    0.73    _{  -0.16   }^{ +0.13   }$&$    0.23    _{  -0.06   }^{ +0.03   }$&$    0.29    _{  -0.10   }^{ +0.10   }$&$    <0.06   $&$    <1.08   $&$    0.15    _{  -0.01   }^{ +0.03   }$\\
4   &$  0.02    _{  -0.02   }^{ +0.14   }$&$    1.65    _{  -0.87   }^{ +0.28   }$&$    0.22    _{  -0.11   }^{ +0.04   }$&$    0.52    _{  -0.20   }^{ +0.09   }$&$    0.22    _{  -0.05   }^{ +0.08   }$&$    0.23    _{  -0.09   }^{ +0.08   }$&$    <0.08   $&$    <1.08   $&$    0.13    _{  -0.01   }^{ +0.02   }$\\
5   &$  0.12    _{  -0.12   }^{ +0.18   }$&$    1.42    _{  -0.58   }^{ +0.28   }$&$    0.23    _{  -0.03   }^{ +0.03   }$&$    0.49    _{  -0.20   }^{ +0.14   }$&$    0.21    _{  -0.06   }^{ +0.09   }$&$    0.10    _{  -0.10   }^{ +0.13   }$&$    <0.15   $&$    <1.12   $&$    0.13    _{  -0.01   }^{ +0.03   }$\\
6   &$  0.01    _{  -0.01   }^{ +0.52   }$&$    2.35    _{  -2.35   }^{ +0.41   }$&$    0.28    _{  -0.14   }^{ +0.10   }$&$    1.05    _{  -0.19   }^{ +0.23   }$&$    0.23    _{  -0.10   }^{ +0.13   }$&$    0.33    _{  -0.21   }^{ +0.25   }$&$    <0.09   $&$    <0.62   $&$    0.16    _{  -0.02   }^{ +0.04   }$\\
\hline
1   &$  <0.37  $&$    <1.82  $&$    <0.14  $&$    0.20    _{  -0.20   }^{ +0.26   }$&$    0.27    _{  -0.09   }^{ +0.16   }$&$    0.24    _{  -0.04   }^{ +0.10   }$&$    <0.22   $&$    <0.36   $&$    0.24    _{  -0.00   }^{ +0.06   }$\\
2   &$  <0.29  $&$    <1.48  $&$    <0.13  $&$    0.50    _{  -0.46   }^{ +0.28   }$&$    0.24    _{  -0.12   }^{ +0.14   }$&$    0.28    _{  -0.05   }^{ +0.13   }$&$    <0.26   $&$    <0.30   $&$    0.28    _{  -0.02   }^{ +0.04   }$\\
3   &$  <0.39  $&$    <1.41  $&$    <0.42  $&$    0.47    _{  -0.37   }^{ +0.48   }$&$    0.37    _{  -0.10   }^{ +0.22   }$&$    0.16    _{  -0.04   }^{ +0.11   }$&$    <0.18   $&$    <0.55   $&$    0.27    _{  -0.01   }^{ +0.05   }$\\
4   &$  <0.48  $&$    <3.08  $&$    <0.48  $&$    0.86    _{  -0.55   }^{ +0.78   }$&$    0.25    _{  -0.25   }^{ +0.30   }$&$    0.30    _{  -0.10   }^{ +0.20   }$&$    <0.34   $&$    <0.27   $&$    0.42    _{  -0.03   }^{ +0.05   }$\\
5   &$  <0.79  $&$    <1.94  $&$    <0.12  $&$    1.19    _{  -0.51   }^{ +0.82   }$&$    0.31    _{  -0.25   }^{ +0.27   }$&$    0.45    _{  -0.13   }^{ +0.23   }$&$    <0.25   $&$    <0.85   $&$    0.44    _{  -0.03   }^{ +0.06   }$\\
6   &$  <1.00  $&$    <5.02  $&$    <0.41  $&$    0.70    _{  -0.36   }^{ +0.29   }$&$    0.25    _{  -0.16   }^{ +0.12   }$&$    0.16    _{  -0.09   }^{ +0.08   }$&$    <0.16   $&$    <0.29   $&$    0.23    _{  -0.03   }^{ +0.03   }$\\

\noalign{\smallskip} \hline \noalign{\smallskip}
\end{tabular}
}
\end{table*}

\begin{figure*}
\includegraphics[width=6cm,clip]{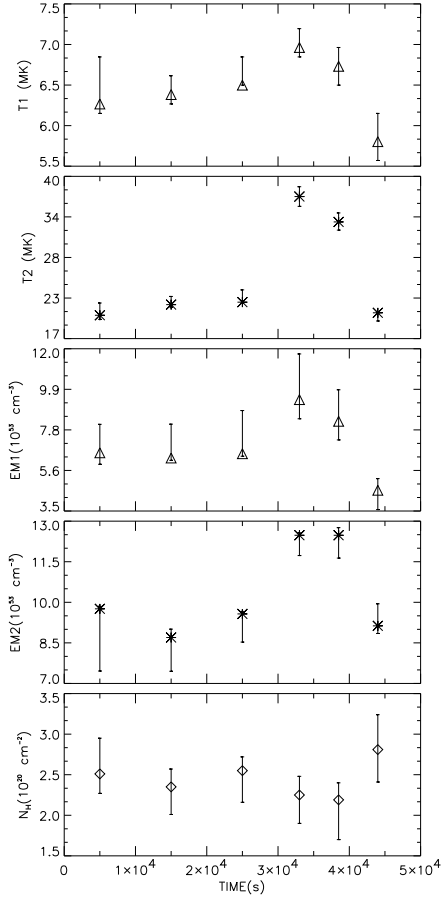}
\caption{Time distribution of the spectral parameters temperature
(top), emission measure (middle) and column density for the cool
component (index 1, ``$\bigtriangleup$'') and hot component (index 2,
``$\ast$''). $N_H$, which is unique for the two components,
is marked with ``$\Diamond$''.}
\end{figure*}

\begin{figure*}
\includegraphics[width=9cm,clip]{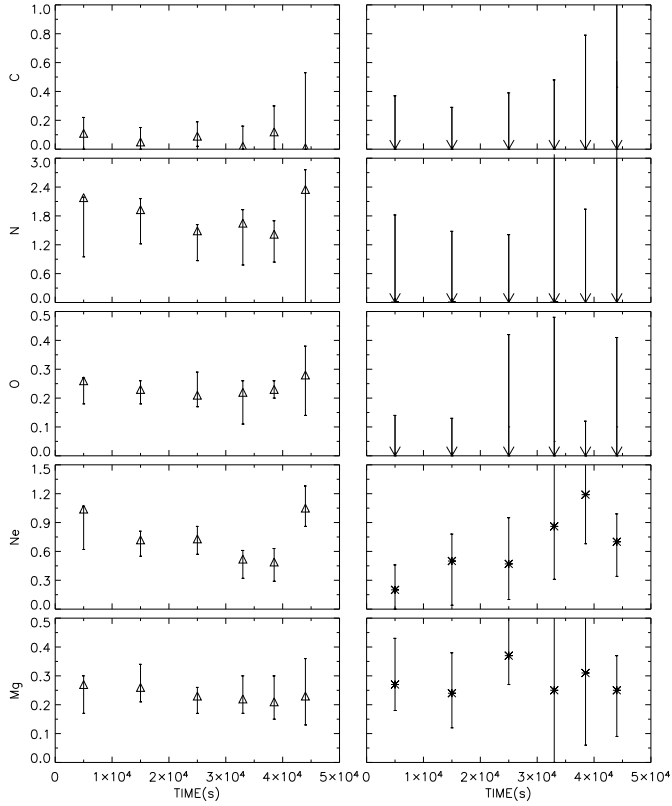}
\caption{Time distribution of the abundances of C, N, O, Ne, Mg for
the cool component (``$\bigtriangleup$'') and hot component
(``$\ast$'').}
\end{figure*}

\begin{figure*}
\includegraphics[width=9cm,clip]{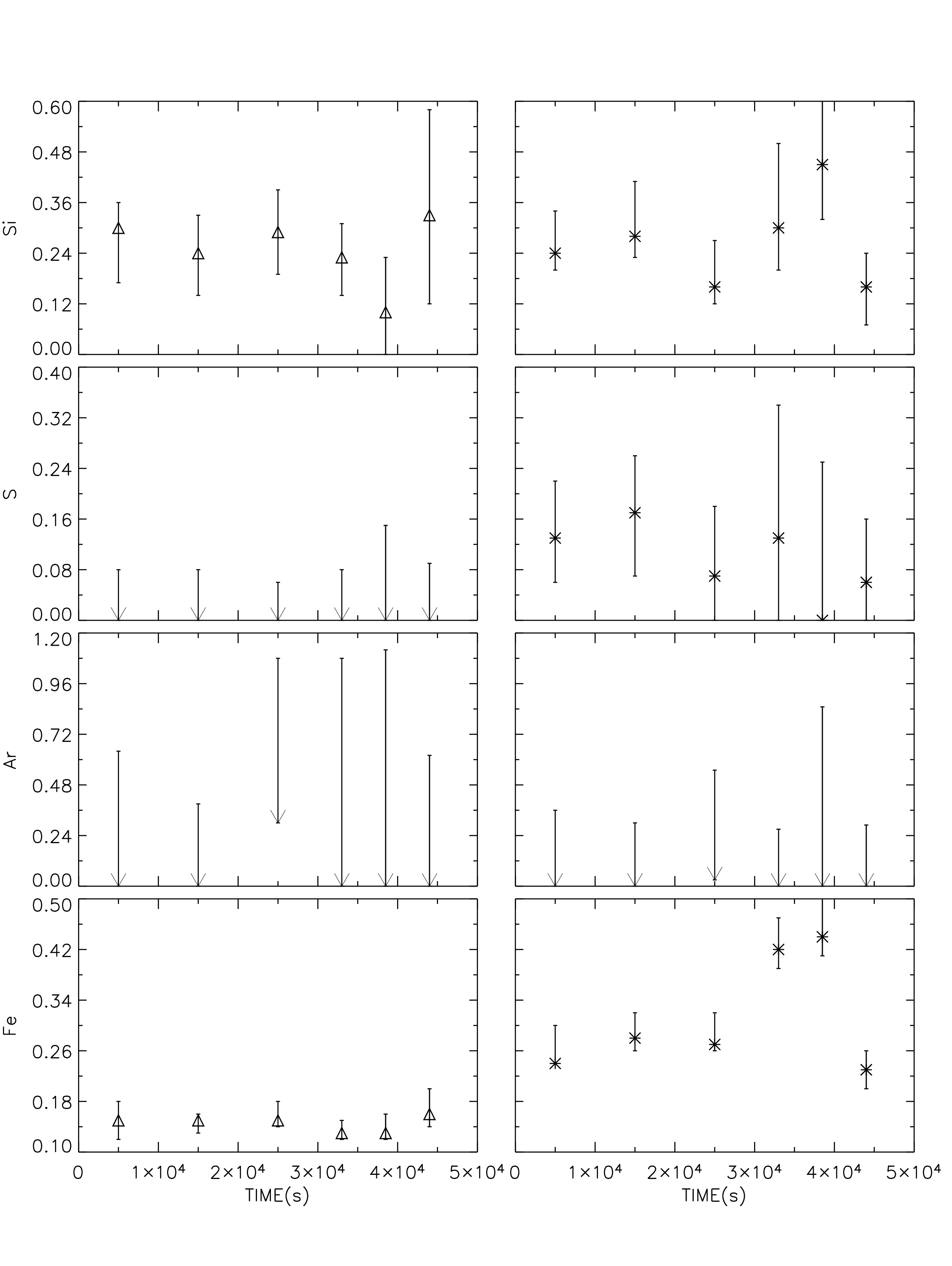}
\caption{Time distribution of the abundances of
Si, S, Ar and Fe for the cool component (``$\bigtriangleup$'')
and hot component (``$\ast$'').}
\end{figure*}

Table 3 demonstrates that the abundance pattern is generally compatible
with that obtained from {\sl ASCA} (Antunes et al. 1994) and {\sl Chandra}
observations (Schmitt \& Ness 2004). An exception is observed for oxygen
abundance of the hot component. But oxygen, as well as carbon and nitrogen
should be fully ionized at these high temperatures so that emission lines
cannot reliably be used to constrain abundances, which means that in our case
the abundances are poorly determined.
The relatively high N abundance and N/C ratio (for the cool component), which
has been detected earlier by Schmitt \& Ness (2002), Drake (2003) and Schmitt
\& Ness (2004), are clearly present in our observation as well. They therefore
support the suggestion made in the papers quoted above that the N
enhancement is related to the CNO-cycle operating in the core of the star.
Furthermore, the active corona of Algol seems to be Fe-poor, which has been
suggested before by Antunes et al. (1994) and Stern et al. (1995). Iron is
likely to be depleted in the corona (Drake 2003).

\begin{table*}
\caption[]{Comparison of coronal elemental abundances from {\sl XMM-Newton}
(this paper), {\sl ASCA} (Antunes et al. 1994) and {Chandra} LETGS
(Schmitt \& Ness 2004).}
\label{obslog}
\scriptsize{
\begin{tabular}{ccccccc}
\noalign{\smallskip} \hline\hline \noalign{\smallskip}
Element&This paper$\ast$&This paper$\ast$&Antunes et al.&Antunes et al.&Antunes et
al.&Schmitt \& Ness\\
&Cool plasma&Hot plasma&Low State&Medium State&High State&\\
\noalign{\smallskip} \hline \noalign{\smallskip}

C    &$  0.07    _{  -0.03   }^{ +0.10   }$&$    0.08    _{  -0.07   }^{ +0.20   }$&$    n.a.
        $& $    n.a.                          $&$    n.a.                            $  &$<0.04$            \\
N    &$  1.84    _{  -0.50   }^{ +0.10   }$&$    0.01    _{  -0.00   }^{ +1.28   }$&$    <0.1                             $& $    <0.1                          $&$    <0.1                            $  &$2.00 $            \\
O    &$  0.24    _{  -0.03   }^{ +0.02   }$&$    0.04    _{  -0.02   }^{ +0.11   }$&$    0.30    _{  -0.04   }^{ +0.04   }$& $  0.31   _{  -0.03   }^{ +0.03   }$&$    0.24   _{  -0.03   }^{ +0.03   }$  &$0.25 $            \\
Ne   &$  0.76    _{  -0.05   }^{ +0.10   }$&$    0.65    _{  -0.17   }^{ +0.22   }$&$    0.76    _{  -0.10   }^{ +0.10   }$& $  1.22   _{  -0.08   }^{ +0.08   }$&$    1.08   _{  -0.08   }^{ +0.08   }$  &$0.95 $            \\
Mg   &$  0.24    _{  -0.03   }^{ +0.03   }$&$    0.28    _{  -0.07   }^{ +0.09   }$&$    0.48    _{  -0.06   }^{ +0.06   }$& $  0.64   _{  -0.05   }^{ +0.05   }$&$    0.47   _{  -0.04   }^{ +0.04   }$  &$0.50 $            \\
Si   &$  0.25    _{  -0.05   }^{ +0.05   }$&$    0.27    _{  -0.07   }^{ +0.09   }$&$    0.43    _{  -0.05   }^{ +0.05   }$& $  0.65   _{  -0.04   }^{ +0.05   }$&$    0.47   _{  -0.03   }^{ +0.03   }$  &$0.45 $            \\
S    &$  0.01    _{  -0.01   }^{ +0.03   }$&$    0.09    _{  -0.03   }^{ +0.06   }$&$    0.29    _{  -0.07   }^{ +0.07   }$& $  0.21   _{  -0.05   }^{ +0.05   }$&$    0.09   _{  -0.04   }^{ +0.04   }$  &$n.a. $            \\
Ar   &$  0.26    _{  -0.12   }^{ +0.23   }$&$    0.14    _{  -0.08   }^{ +0.12   }$&$    0.10    _{  -0.10   }^{ +0.10   }$& $  0.14   _{  -0.12   }^{ +0.12   }$&$    0.10   _{  -0.11   }^{ +0.11   }$  &$n.a. $            \\
Fe   &$  0.14    _{  -0.01   }^{ +0.01   }$&$    0.31    _{  -0.01   }^{ +0.02   }$&$    0.30    _{  -0.01   }^{ +0.01   }$& $  0.37   _{  -0.02   }^{ +0.02   }$&$    0.32   _{  -0.01   }^{ +0.01   }$  &$0.20 $            \\

\noalign{\smallskip} \hline \noalign{\smallskip}
\end{tabular}
}

$\ast$ Average of best-fit values of the 6 spectra \\
\end{table*}

Comparing the abundances of the cool and the hot component (c.f. Table 2, Fig.
3 \& Fig. 4) we note that
apart from perhaps the light elements C, N, O there appear to be no significant
differences for Ne, Mg and Si. Because of the large uncertainties also the
abundances of S and Ar are not really accessible, but upper limits can be
provided. A clear difference, however, exists for Fe (c.f. Fig. 6).
In fact, F-tests show that we cannot rule out that the abundances of the
cool and the hot component are identical but with the exception of Fe. Fig. 6
shows the best-fit and the corresponding error contours for the Fe abundances
of the cool and hot component of spectrum \#1, i.e. during the quiescent phase.
Up to the 4$\sigma$ level the abundances differ from each other. For the best
fit the abundances are Fe(cool)$\sim$ 0.14 and  Fe(hot)$\sim$ 0.28. This result
does not only hold for spectrum \#1 but for the entire quiescent phase including
the post-flare phase covered by spectrum \#6. From the {\sl ROSAT} PSPC observation
(Ottmann 1994) it was concluded that the low and high temperature components
are rather co-spatial with respect to the stellar surface. We therefore
speculate that the cool and hot components might come from different layers or
that the hot regions are much smaller in size than the radius of the K star.

\begin{figure*}
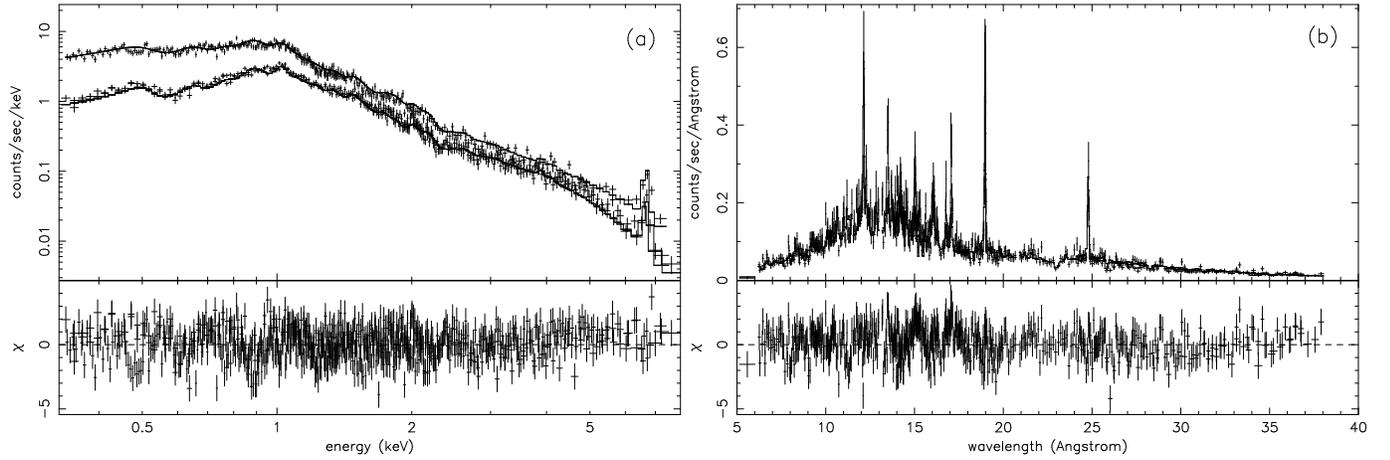

\includegraphics[width=6cm,angle=270,clip]{ms575fig5a.ps}
\includegraphics[width=6cm,angle=270,clip]{ms575fig5b.ps}
\caption{Two-temperature thermal plasma model fits to
 spectrum \#4. Panel a is for EPIC, where the top curve is for
 the PN and the bottom curve is for MOS1 and MOS2, while b
 is for RGS1 and RGS2. The x-axis for RGS is wavelength instead of
 energy and y-axis is not plotted in log scale, to make it much easy
 to recognize the emission lines.
 The EPIC spectra show the presence of an Fe 6.7 keV line.}
\end{figure*}

\begin{figure*}
\includegraphics[width=6cm,angle=270,clip]{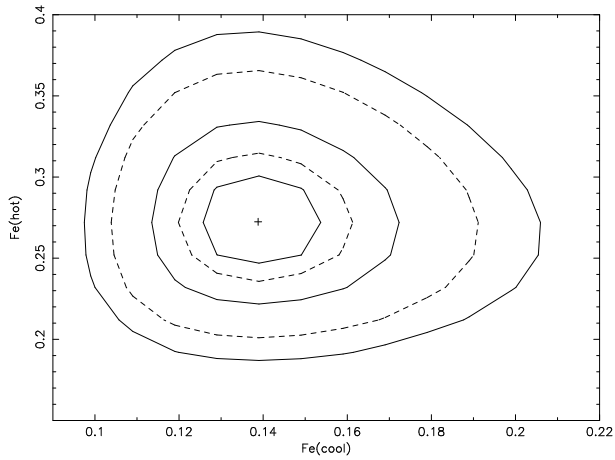}
\caption{Fe abundance of the hot component versus the Fe abundance of the
cool component obtained from spectrum \#1 of the quiescent phase.
The best-fit values are marked by a cross; the contours correspond
to errors of 1$\sigma$ to 5$\sigma$ stepped by 1$\sigma$.}
\end{figure*}

We admit that the two components are not likely to represent physically distinct
temperatures. A temperature gradient in the corona might exist, or the two
temperatures are just a short-cut expression of a temperature dependent
emission measure.
In order to check whether the different abundances of the cool and hot
components we obtained above are real or only the consequence of an oversimplified
temperature model, we also fit the spectra with a three-temperature vmeka model.
Similar to what we have done in the two-temperature vmeka model fitting, we first
set the three iron abundances as free parameters and then make the three components
to share the same iron abundance. The fitting results are given in Table 4. Ftest
shows that for spectra \#1 - 5 the probabilities that the three components have the
same abundance are lower than $5\times 10^{-5}$, similar to those of the two-temperature
model fitting. The only exception is spectrum \#6, for which the corresponding probability
is 0.36, and one possible reason is that spectrum \#6 contains fewer counts than the
other spectra. In Fig. 8, as an example, we show the comparison of the three-temperature
spectral fittings to spectrum \#5, with the three iron abundance equal to or different from
each other. The apparent residual bump around 6.7 keV in the left panel also implies that
the hot component has a higher iron abundance.

\begin{table*}
\caption[]{Main spectral properties of Algol (3T VMEKAL model)}
\label{obslog}
\scriptsize{
\begin{tabular}{cccccccccc}
\noalign{\smallskip} \hline\hline \noalign{\smallskip}
No.&T1&T2&T3&\multicolumn{3}{c}{Fe}&$\chi^2_{\sl red}$&DOF$\ast$&F-test$\dagger$\\
&($10^6$K)&($10^6$K)&($10^6$K)& Fe1&Fe2 &Fe3 & & &probability\\
\noalign{\smallskip} \hline \noalign{\smallskip}

1   &$  4.76    _{  -0.35   }^{+    0.23    }$&$    8.24    _{  -0.46   }^{+    0.58    }$&$    23.91   _{  -1.04   }^{+    1.16    }$&$    0.17    _{  -0.02   }^{+    0.04    }$&$    0.18    _{  -0.03   }^{+    0.06    }$&$    0.31    _{  -0.04   }^{+    0.04    }$& 1.45    &   1982    &  \\
2   &$  4.76    _{  -0.35   }^{+    0.23    }$&$    8.24    _{  -0.35   }^{+    0.58    }$&$    25.53   _{  -1.28   }^{+    1.39    }$&$    0.17    _{  -0.03   }^{+    0.05    }$&$    0.18    _{  -0.04   }^{+    0.06    }$&$    0.35    _{  -0.04   }^{+    0.05    }$& 1.50    &   2003    &  \\
3   &$  4.64    _{  -0.35   }^{+    0.23    }$&$    8.12    _{  -0.35   }^{+    0.46    }$&$    26.92   _{  -1.28   }^{+    1.39    }$&$    0.25    _{  -0.06   }^{+    0.12    }$&$    0.15    _{  -0.03   }^{+    0.03    }$&$    0.36    _{  -0.05   }^{+    0.04    }$& 1.44    &   2054    &  \\
4   &$  4.99    _{  -0.35   }^{+    0.23    }$&$    9.40    _{  -0.93   }^{+    0.70    }$&$    40.39   _{  -2.09   }^{+    2.79    }$&$    0.15    _{  -0.02   }^{+    0.04    }$&$    0.22    _{  -0.07   }^{+    0.09    }$&$    0.41    _{  -0.06   }^{+    0.05    }$& 1.37    &   1329    &  \\
5   &$  4.99    _{  -0.93   }^{+    0.35    }$&$    9.05    _{  -1.28   }^{+    0.70    }$&$    34.93   _{  -1.04   }^{+    2.79    }$&$    0.17    _{  -0.03   }^{+    0.14    }$&$    0.22    _{  -0.09   }^{+    0.12    }$&$    0.45    _{  -0.06   }^{+    0.07    }$& 1.36    &   1139    &  \\
6   &$  4.76    _{  -0.12   }^{+    0.35    }$&$    8.36    _{  -0.23   }^{+    1.16    }$&$    23.09   _{  -1.97   }^{+    0.81    }$&$    0.20    _{  -0.05   }^{+    0.02    }$&$    0.27    _{  -0.03   }^{+    1.05    }$&$    0.26    _{  -0.08   }^{+    0.02    }$& 1.52    &   960 &  \\
\hline
1   &$  4.87    _{  -0.35   }^{+    0.23    }$&$    8.36    _{  -0.35   }^{+    0.35    }$&$    22.63   _{  -0.93   }^{+    0.81    }$& \multicolumn{3}{c}{$    0.22    _{  -0.02   }^{+    0.02    }$}&    1.47    &   1984    &   $5.12\times10^{-5}$\\
2   &$  4.99    _{  -0.35   }^{+    0.12    }$&$    8.59    _{  -0.46   }^{+    0.35    }$&$    23.56   _{  -0.93   }^{+    1.04    }$& \multicolumn{3}{c}{$    0.23    _{  -0.02   }^{+    0.03    }$}&    1.52    &   2005    &   $3.43\times10^{-8}$\\
3   &$  4.99    _{  -0.23   }^{+    0.23    }$&$    8.82    _{  -0.23   }^{+    0.46    }$&$    24.60   _{  -1.04   }^{+    1.39    }$& \multicolumn{3}{c}{$    0.24    _{  -0.02   }^{+    0.02    }$}&    1.47    &   2056    &   $2.29\times10^{-5}$\\
4   &$  5.22    _{  -0.12   }^{+    0.35    }$&$    9.75    _{  -0.23   }^{+    0.70    }$&$    40.39   _{  -1.28   }^{+    5.45    }$& \multicolumn{3}{c}{$    0.26    _{  -0.05   }^{+    0.02    }$}&    1.40    &   1331    &   $1.98\times10^{-7}$\\
5   &$  5.34    _{  -0.35   }^{+    0.23    }$&$    9.52    _{  -0.70   }^{+    0.46    }$&$    34.23   _{  -1.86   }^{+    2.32    }$& \multicolumn{3}{c}{$    0.31    _{  -0.05   }^{+    0.05    }$}&    1.39    &   1141    &   $5.53\times10^{-7}$\\
6   &$  4.64    _{  -0.35   }^{+    0.35    }$&$    8.24    _{  -0.46   }^{+    0.35    }$&$    23.09   _{  -0.93   }^{+    1.04    }$& \multicolumn{3}{c}{$    0.24    _{  -0.03   }^{+    0.03    }$}&    1.52    &   962     &   $3.60\times10^{-1}$\\

\noalign{\smallskip} \hline \noalign{\smallskip}
\end{tabular}
}

$\ast$ degrees of freedom.\\
$\dagger$ F-test probability for adopting identical iron abundance for the three components.\\

\end{table*}

(IV) The abundance of Fe and possibly those of Ne and Si of the hot component
appear to change during the flare (c.f. Table 2). The Fe abundance increased
from 0.25 times the solar value in the quiescent state to about 0.5 during the flare.
To put this in a more quantitative context we carried out an F-test and assess
whether the elemental abundances are identical for the quiescent and the flare-only
emission. This hypothesis can indeed be accepted but only if the abundances of
Fe between the hot components differ. The cool component is consistent with one
set of Fe abundance for the quiescent and the flare-only phase.
Fig. 7 shows the best fit  and the error contours for the Fe abundance of the
hot versus the cool component for a flare spectrum. The Fe-abundances are
inconsistent with each other well beyond the 5$\sigma$ level.
Whereas the Fe abundance of the cool component stays at $\sim$0.14, the best fit
for the Fe abundance of the hot component is now at $\sim$0.44, which means an
increase by a factor of $\sim$1.6 between the quiescent and the flare phase.
This kind of behavior of the Fe abundance still holds for the
 three-temperature model, i,e.,  the Fe abundances for the cool and
middle component remain unchanged, while that of the hot component increases during
the flare, although not as much as that in the two-temperature model.
This kind of increase for Fe abundance during the flare have also been detected in
CN Leonis (Liefke et al. 2010)

\begin{figure*}
\includegraphics[width=6cm,angle=270,clip]{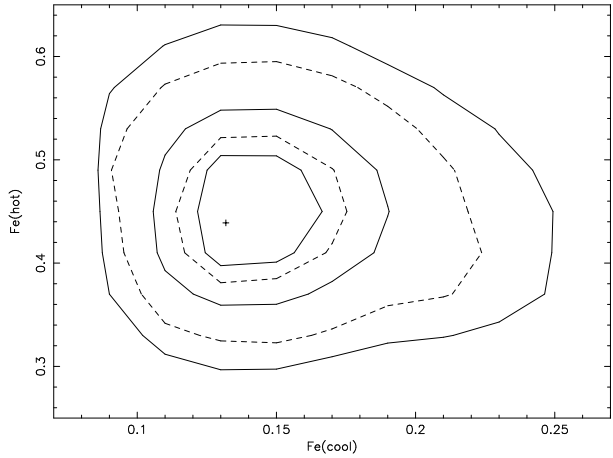}
\caption{Fe abundance of the hot component versus the Fe abundance of the
cool component obtained from spectrum \#5 of the flare phase.
The best-fit values are marked by a cross; the contours correspond
to errors of 1$\sigma$ to 5$\sigma$ stepped by 1$\sigma$.}
\end{figure*}

\begin{figure*}
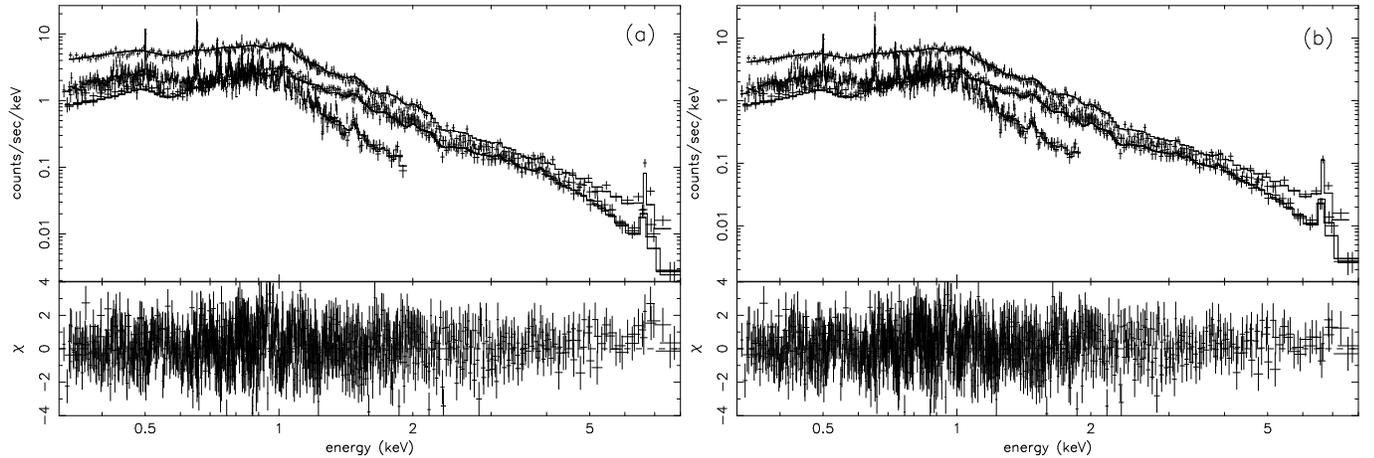

\includegraphics[width=6cm,angle=270,clip]{ms575fig8a.ps}
\includegraphics[width=6cm,angle=270,clip]{ms575fig8b.ps}
\caption{Three-temperature thermal plasma model fits to
 spectrum \#5. Panel a is for a global abundances of all the elements for the
 three component, while b is for only Fe relaxed.}
\end{figure*}

This Fe abundance increase during a flare may be
attributed to chromospheric evaporation (Favata \& Micela 2003). The
abundances of Ne and Si for the hot component may show a similar behavior
although with much larger uncertainty. This could further support the idea of
chromospheric evaporation. The elemental abundances of
the cool component remain nearly unchanged during the entire observation,
which is additional support for the hot component dominating the flare.

\subsection{The flare-only spectra}
The spectra of the flare-only emission can be well fitted with a
one-temperature MEKAL model.  Table 5 gives the fitted parameters
and Fig. 9 shows the model fit to the flare-only spectrum \#2 (corresponding
to the overall spectrum \#5 ). The temperature of the flare emission
is around 6.2$\times$10$^7$ K and it shows no
detectable variations during the flare.
 Obviously the flare temperature is much higher than the
temperature of the hot component of the quiescent state.
The elemental abundances are
comparable to those of the hot component and they show no variation either
(see Table 4). The best-fit $N_H$ of the flare emission is consistent
with no absorption at all for each of the two spectra. Taking into account
the uncertainties it is still significantly lower than the $N_H$ values associated
with the quiescent emission, which is consistent with the results of the
{\sl ROSAT} observation (Ottmann 1994; Ottmann \& Schmitt 1996).
We then used the F-test to examine whether the $N_H$ difference between
the flare emission and the quiescent
emission is real or just a statistical fluctuation. This test is
based on the comparison of the $\chi^2$ values and the degrees of freedom of the model fits
to the flare-only spectra with $N_H$ as a free parameter against
$N_H$ being fixed to the mean $N_H$ of the quiescent emission (2.5
$\times10^{20}$ cm$^{-2}$). As shown by the values listed in Table 4,
the $N_H$ of the flare emission is lower than that of  the quiescent
emission at a significance level greater than 99.5\%.
A similar behavior is suggested by the ROSAT observation with an
overall $N_H$ of about 5$\times10^{19}$ cm$^{-2}$ (Ottmann 1994), while
the flare-only $N_H$ is around 1$\times10^{19}$ cm$^{-2}$ (Ottmann \&
Schmitt 1996).

\begin{table*}
\caption[]{Best fit results for flare-only spectra} \label{obslog}
\label{obslog}
\scriptsize{
\begin{tabular}{cccccccccc}
\noalign{\smallskip} \hline\hline \noalign{\smallskip}

No.&t&$N_H$&T&Z&EM&flux&$\chi^2_{\sl red}$&DOF&F-test\\
&(s)&($\frac
{10^{20}}{cm^2}$)&($10^6$K)&($Z_{\odot}$)&($\frac{10^{53}}{cm^3}$)&($\frac
{10^{-11}erg}{{cm^2}s}$)&&&probability\\

\noalign{\smallskip} \hline \noalign{\smallskip}

1&6000&$<0.60(2\sigma)$&$63.6_{-4.8}^{+4.4}$&$0.55_{-0.12}^{+0.12}$&$5.33_{-0.15}^{+0.21}$
&9.24&1.11&373&$9.50\times10^{-6}$\\
2&5000&$<1.58(2\sigma)$&$60.1_{-5.4}^{+4.1}$&$0.64_{-0.15}^{+0.16}$&$4.56_{-0.23}^{+0.25}$
&7.94&1.02&264&$5.71\times10^{-3}$\\

\noalign{\smallskip} \hline \noalign{\smallskip}
\end{tabular}
}
\end{table*}

\begin{figure*}
\includegraphics[width=6cm,angle=270,clip]{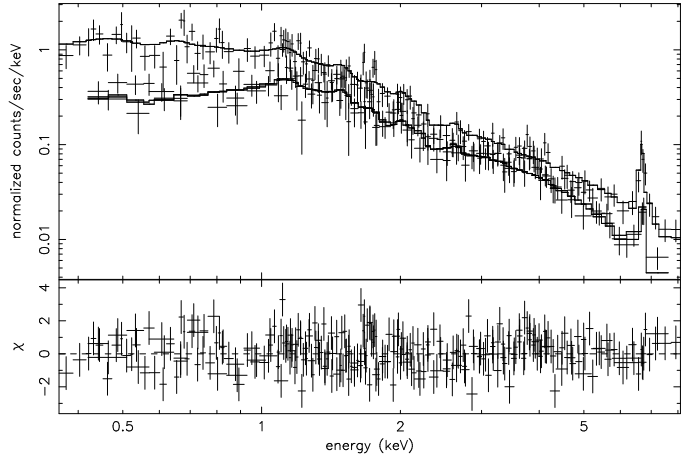}
\caption{The flare-only spectrum \#2. The upper curve is for the
PN and the lower curve is for MOS1 and MOS2.}
\end{figure*}


In order to check whether the above results are intrinsic properties
of the flare-only emission or are related to the particular model for the background
spectrum, we also tried  the quiescent spectrum in time interval 2 as
the background spectrum to create the flare-only spectra. These spectra are
fitted with the same MEKAL model, and the results are well consistent with
the previously obtained ones. In fact, this is not surprising given that
the parameters of the overall spectra \# 1 and \# 2 are consistent with each
other within the error bars. We therefore conclude that values for the low absorption
column density, the high temperature and the high metallicity of the flare-only emission
are robust.

\section{Cool absorbing gas surrounding the K star}
The mean $N_H$ value of the quiescent emission derived in this paper is
about 2.5$\times10^{20}$ cm$^{-2}$, which is at least
two orders of magnitude higher than the interstellar  H{\small I} + H$_2$
column density reported by Welsh et al. (1990).
The low energy calibration of {\sl XMM} may affect our results, but not
as much as two orders of magnitude as stated in $\S$ 3.1.
Stern et al. (1995) estimated the $N_H$ to be 2$\times10^{18}$ cm$^{-2}$
using EUVE observation, which is consistent with that of Welsh et al. (1990).
The $N_H$ of 5$\times10^{19}$ cm$^{-2}$ obtained in the {\sl ROSAT} PSPC observation is
lower than our result (Ottmann 1994), but still 20 times higher than the
ISM column density. However, since Welsh et al. (1990) inferred the ISM
column density using the Na line absorption by the continuum
of the B type primary that is X-ray dark, and since the $N_H$ values
obtained in X-ray observations correspond to the
absorbing column densities of the X-ray emitting K type secondary, the higher
$N_H$ values thus imply that there exists cool gas just surrounding the K star.
Furthermore, the result of Stern et al. (1995) was based on the He/H ratio and
with no ionized He. This would be the case for our model of ionized local
absorbing material, which would make the EUV transparent but not to X-rays.
Because no orbital (and so rotational) modulation of $N_H$
has been detected by either {\sl ROSAT} (Ottmann 1994) or {\sl XMM-Newton},
the cool gas should be distributed quite homogeneously relative to the
longitude of the K star surface.  The $N_H$ difference between
the {\sl ROSAT} and {\sl XMM-Newton} observations might further
suggest that the amount of cool gas surrounding the K star changes
over a long term.

The location and configuration of this cold matter is
unclear. However, in the framework of the two polar lobe model
(Favata et al. 2000, Retter et al. 2005, Peterson et al. 2010) for the site of the
quiescent corona, one could imagine that the absorption column
density is highest in or around those lobes, where the quiescent corona is
dominating. It is reduced towards the equatorial regions, at which
the certainly non-polar flare of this {\sl XMM-Newton} observation
is located.

Such a configuration would also explain the observation of the 133
\AA \ line. In the polar regions, i.e. the lobes, it is likely to be
completely absorbed, but it would become observable if only a small
fraction of coronal emission would also exist in sub-polar regions
of low absorption. That the corona stretches down from the polar
regions has been shown by Schmitt et al. (2003) discussing the
location of the quiescent coronal emission of just these {\sl
XMM-Newton} observations.

\section{Summary}

We have analyzed an {\sl XMM-Newton} observation of the eclipsing
binary Algol, in which the interval of the secondary optical minimum
and some preceding section in the orbit were covered. During the
eclipse of the X-ray bright K star a flare occurred, so that we can
study both the quiescent corona and an eclipsed flare. We joined the
data of all four X-ray instruments on board of {\sl XMM-Newton},
which give spectral data at high resolution with the RGSs from 0.3 -
2 keV and medium resolution data with the EPIC cameras from 0.3 - 10
keV.

Satisfactory fits have been obtained over the entire energy band
using the two-temperature VMEKAL model for the overall spectrum. 
We cannot rule out that the elemental abundances of the spectrum's
low temperature and high temperature components are identical, but the abundance of Fe is clearly
different at the 4$\sigma$ level for the quiescent phase and significantly more
than 5$\sigma$  for the flare phase, indicating that the high temperature  and low temperature plasma
components are not well mixed. We also observed a significant Fe abundance increase in the hot
component by a factor of $\sim$ 1.6 during the flare, which supports the idea
of chromospheric evaporation.

The fits to the $N_H$ column density reveal values around
2.5$\times10^{20}$ cm$^{-2}$, for the quiescent corona which is far
in excess of the interstellar column density. On the other hand, the
fits to the flare-only emission are consistent with no absorption
exceeding the interstellar value. We propose that the line of sight
column density across Algol is not uniform. It is sufficiently high
towards the polar regions, and reduced towards the equatorial regions.

\normalem
\begin{acknowledgements}
We thank the anonymous referee for the very helpful comments.
This project is supported by the National Natural Science Foundation of
China under Nos. 10903007 and 10778716. This work is based
on observations with XMM-Newton, an ESA science mission with instruments
and contributions directly funded by ESA member states and the USA (NASA).
\end{acknowledgements}

\end{document}